
\magnification=\magstep1
\def\pn{\par\noindent}
\def\sj{\vskip 0.9cm \noindent}
\def\vL#1{\bar{\Lambda}_{#1}}
\def\m@th{\mathsurround=0pt}
\def\Fsquare(#1,#2){
\hbox{\vrule$\hskip-0.4pt\vcenter to #1{\normalbaselines\m@th
\hrule\vfil\hbox to #1{\hfill$#2$\hfill}\vfil\hrule}$\hskip-0.4pt
\vrule}}

\def\Addsquare(#1,#2){\hbox{$
	\dimen1=#1 \advance\dimen1 by -0.8pt
	\vcenter to #1{\hrule height0.4pt depth0.0pt%
	\hbox to #1{%
	\vbox to \dimen1{\vss%
	\hbox to \dimen1{\hss$~#2~$\hss}%
	\vss}%
	\vrule width0.4pt}%
	\hrule height0.4pt depth0.0pt}$}}

\def\Vtwobox(#1,#2){%
	\normalbaselines\m@th\offinterlineskip
	\vcenter{\hbox{\Fsquare(0.5cm,#1)}
	      \vskip-0.4pt
	      \hbox{\Fsquare(0.5cm,#2)}}}

\def\naga{%
	\hbox{$\vcenter to 0.5cm{\normalbaselines\m@th
	\hrule\vfil\hbox to 1.2cm{\hfill$\cdots$\hfill}\vfil\hrule}$}}

\def\Yang{1}
\def\Bax{2}
\def\Jimbo{3}
\def\Baz{4}
\def\KSR{5}
\def\OW{6}
\def\RW{7}
\def\KirR{8}
\def\Drin{9}
\def\ChPrcmp{10}
\def\KNSo{11}
\def\AKC{12}
\def\Resan{13}
\def\BR{14}
\def\Og{15}
\def\AKG{16}
\def\KSG{17}
\def\ChPra{18}
\def\analyR{19}
\def\MacC{20}
\def\AKBAE{21}
\def\AKJS{22}
\def\rfund{4}
\def\vacfund{5}
\def\dresses{6}
\def\bae{7}
\def\boxes{8}
\def\restau{9}
\def\tttot{10}
\def\twoboxes{11}
\def\newvac{12}
\def\tsys{13}
\def\yangch{14}
\def\soltau{15}
 \vskip 2cm
\centerline{\bf Fusion $U_q(G^{(1)}_2)$ vertex models}\par
\centerline{\bf and }\par
\centerline{\bf analytic Bethe ans{\"a}tze}\par
 \vskip 2cm
\centerline{ {\bf Junji Suzuki}
\footnote*{e-mail address:jsuzuki@tansei.cc.u-tokyo.ac.jp}}
\centerline{\sl Institute of Physics, College of Arts and Sciences}\par
\centerline{\sl University of Tokyo}\par
\centerline{\sl Komaba 3-8-1, Meguro-ku, Tokyo}\par
\centerline{\sl Japan}\par
 \vskip 3cm
\noindent We introduce fusion $U_q(G^{(1)}_2)$ vertex models
 related to fundamental representations.
The eigenvalues of their row to row transfer matrices are derived through
analytic Bethe ans{\"a}tze.
By combining these results with
our previous studies on functional relations among
transfer matrices(the $T$-system), we conjecture  explicit eigenvalues for
a wide class of fusion models. These results can be neatly
expressed in terms of a Yangian analogue of the Young tableaux.

\vfill\eject
\noindent\S 1 {\bf Introduction}\pn
Recent years, the quantum group symmetry has been playing
major roles in various branches of physics.
 Yang Baxter equation is  the most fundamental relation
in these studies[\Yang, \Bax]:
$$
R_{W_2 W_3}(u) R_{W_1 W_3}(u+v) R_{W_1 W_2}(v) =
R_{W_1 W_2}(v) R_{W_1 W_3}(u+v) R_{W_2 W_3}(u).
\eqno{(1)}
$$
Here $R-$matrix, $R_{W_1 W_2}(u)$ denotes the operator acting upon
a triplex of spaces $W_1 \otimes W_2 \otimes W_3$ as
$R_{W_1 W_2}(u) \otimes 1$.
Viewed as a two dimensional(2D) vertex model, $R_{W_1 W_2}(u)$ represents
the vertex weights acting on the auxiliary (horizontal) space $W_1$
and the quantum (vertical) space $W_2$.
 Jimbo and Bazhanov[\Jimbo,\Baz]
 found a systematic way of constructing solution
to eq.(1).
Let $V_{\Lambda} $ be a highest weight module of a
Lie algebra $g$ with the highest weight $\Lambda$.
Once a solution for $W_1=W_2=W_3=V_{\vL{1}}$ is known,
 it is in principle possible to
obtain $R-$matrices acting on spaces $W_1, W_2$ and $ W_3$
other than $V_{\vL{1}}$ and not
necessary identical[\KSR].
This procedure is quite general and called the fusion.\pn
The fusion space, however, is not irreducible as the $g-$ module
in general[\OW,\RW,\KirR].
%
 For $V_{m \vL{a}}$,  we have the corresponding
 fusion space $W^{(a)}_m$, that is, the irreducible finite dimensional
representation of quantum affine algebra $U_q(\hat{g})$[\Jimbo, \Drin].
It is expected that the irreducible finite dimensional
representation of quantum affine algebra for generic $q$
is smoothly connected to that of Yangian $q \rightarrow 1$
(for $g=sl_2$, see [\ChPrcmp]).
 In this sense, we call $W^{(a)}_m$ Yangian space in the
following.
Then the decomposition of $W^{(a)}_m$ as the $g-$ module reads,
$$
W^{(a)}_m |_{g}= \oplus_{\{n\}} Z(m,\{ n \},\Lambda') V_{\Lambda'}
\eqno{(2)}
$$
where $\Lambda' = m \vL{a} -\sum n_b \alpha_b$ and
$Z(m,\{ n \},\Lambda')$ is a known numerical factor given in [\KirR].

In this communication, we work on fusion  $U_q(G^{(1)}_2)$
vertex models. The motivation comes from our previous study[\KNSo] on the
functional identities among row to row transfer matrices(referred to as
the $T$-system in the following).
Let  the quantum and the auxiliary space be
$W^{(p)}_s \phantom{}^{\otimes N}$ and $W^{(a)}_m$, respectively.
Here $a,p=1, \cdots,\hbox{rank } g$ and $m, s$ denote arbitrary
integers. In the followings we fix $p,s$ for these meanings.
We define a transfer matrix $T^{(a)}_m(u)$ by
$$
T^{(a)}_m(u) = \hbox{tr}_{W^{(a)}_m} R_{W^{(a)}_m, W^{(p)}_s} (u-w_1)
\cdots  R_{W^{(a)}_m, W^{(p)}_s} (u-w_N).
\eqno{(3)}
$$
The $T-$ system states the relations among $T^{(a)}_m(u)$ for
various $a, m$, however, with identical $W^{(p)}_s$.
( See eq.(\tsys) for $U_q(G^{(1)}_2)$).
In ref[\KNSo], it is shown that the $T$-system
Yang-Baxterizes the relations between Yangian characters $Q^{(a)}_m$.
Moreover, it has
been conjectured that any $T^{(a)}_m(u)$ and its eigenvalue
$\tau^{(a)}_m(u)$
 can be written as a polynomial of fundamental ones,
$a=1,\cdots, \hbox{rank }g, m=1$,
irrespective of $W^{(p)}_s$.
Thus, by finding explicit forms for elementary ones, we
can obtain those for  arbitrary $a, m$.
 This program has been recently completed for $U_q(C^{(1)}_2)$ model[\AKC].
For $U_q(G^{(1)}_2)$ model, we have also two fundamental transfer
matrices $T^{(1)}_1(u)$
and $T^{(2)}_1(u)$.  Their explicit eigenvalues for $s=1$
are conjectured in this report via analytic Bethe ans{\"a}tze[\Resan].
We find it convenient to introduce a Yangian analogue of the Young
tableaux in presenting these results.
 A conjecture on the explicit form for $\tau^{(1)}_m(u)$,
$m$ arbitrary integer, is proposed in terms of tableaux
generated by {\it quite simple rules}.
 Such tableaux are first introduced by Bazhanov and Reshetikhin
[\BR] for $s\ell_n$ models.
Their tableaux coincide with the semi-standard tableaux.
We believe that this coincidence is not accidental but universal.
That is, the tableaux representing the eigenvalues of transfer
matrices might be deserved as Yangian basis.
Then the simple rules we find here imply that the Yangian
theory is, though complicated in disguise, quite natural object.
%
\sj
\S 2  {\bf Fundamental Models}\pn
We fix notations. Let $\alpha_1, \alpha_2$ be simple roots.
We introduce the bilinear form $(*|*)$ with normalization
$(\alpha_1|\alpha_1)=3(\alpha_2|\alpha_2)=2$ and $(\alpha_1|\alpha_2)=-1$.
Fundamental weights $\vL{1}, \vL{2}$ are given by
$ \vL{1}=2\alpha_1+3\alpha_2, \vL{2}=\alpha_1 +2\alpha_2$.
Let $\epsilon_{-3}=\alpha_1 +2 \alpha_2, \epsilon_{-2}=\alpha_1+\alpha_2,
\epsilon_{-1}=\alpha_2, \epsilon_0=0$ and
$\epsilon_i=-\epsilon_{-i},(i=1,2,3)$.
The corresponding weight vectors are denoted by $v_{i}, i=1,\cdots, 7$.
These 7 vectors form basis for the highest weight module $V_{\vL{2}}$
 of $G_2$ with the highest weight vector $\vL{2}$.\pn
 $[u]= {{(q^u-q^{-u})}\over{(q-q^{-1})}}$ where $q \not=$ root of unity
 is the deformation parameter.\pn
We first describe the most fundamental vertex model obtained by
 Ogievetsky($q=1$)[\Og] and Kuniba($q$ general)[\AKG]. \pn
model 1. $W_1=W_2=W^{(2)}_1=V_{\vL{2}}$ \pn
Let the above-mentioned 7 vectors be physical variables assigned to each bonds.
At each vertex, the conservation of weights is imposed:
$\epsilon_i+\epsilon_j=\epsilon_k+\epsilon_{\ell}$. This "ice rule" results
175 possible vertex configurations.
Let $P$ be a permutation operator: $P(u\otimes v) =v \otimes u$.
{}From Shur's lemma,
$PR_{V_{\vL{2}},V_{\vL{2}}}(u)$ can be represented using the operators
$P_{2 V_{\vL{2}}} (P_{V_{\vL{1}}}, P_{V_{\vL{2}}},
P_{V_{0}})$ which project  $ V_{\vL{2}} \otimes V_{\vL{2}}$
to $V_{2 \vL{2}}$
($V_{\vL{1}}, V_{\vL{2}}, V_{0})$ , respectively),
$$
PR_{V_{\vL{2}},V_{\vL{2}}}(u)=
\sum_{\Lambda=2\vL{2},\vL{1},\vL{2},0} \rho_{\Lambda}(u) P_{\Lambda}
\eqno{(\rfund)}
$$
where
$$
[4][6]\rho_{\Lambda}(u) =
\cases{
[1+u][4+u][6+u]& $ \Lambda=2\vL{2} $ \cr
[1-u][4+u][6+u]& $ \Lambda= \vL{1} $ \cr
[1+u][4-u][6+u]& $ \Lambda= \vL{2} $ \cr
[1-u][4+u][6-u]& $ \Lambda=0 $. \cr
}
$$
Explicit expressions for projectors $ P_{\Lambda}$ can be found in ref[\AKG].
Let us remark that RSOS counterpart is solved in [\KSG].\pn
{}From the above decomposition, one sees that
$PR_{V_{\vL{2}},V_{\vL{2}}}(u)$ possesses singular
points at $u=1$ etc.
The Yangian space $W^{(1)}_1$ is related to the singularity at $u=1$.
As  the $G_2$ module,  15 dimensional space $W^{(1)}_1$
decomposes as $V_{\vL{1}}
 \oplus V_{0}$.
The basis for the latter two modules in terms of $v_{i} (i=-3,\cdots, 3)$
are obtained in ref.[\AKG].
 There, three vectors
$v^{(\vL{1})}_7, v^{(\vL{1})}_8$ and $v^{(0)}_1$ having zero
weight are constructed explicitly.
  In the module $W^{(1)}_1$, these vectors are, after some modifications,
 constituents of a null space. \pn
$W^{(1)}_1$ and $W^{(2)}_1$ deserve Yangian analogue of
the spaces of fundamental representations[\ChPra].
In the next section, we deal with the eigenvalue problems for
transfer matrices of which the quantum and the auxiliary spaces are
either of $W^{(1)}_1$ or $W^{(2)}_1$. \pn
For convenience, we call,\pn
model 2. $W_1=W^{(1)}_1, W_2=W^{(2)}_1$ \pn
model 3. $W_1=W^{(2)}_1, W_2=W^{(1)}_1$ \pn
model 4. $W_1=W_2=W^{(1)}_1$. \pn
%
%
\sj
\S 3 {\bf Eigenvalues for transfer matrices}\pn
Let us address the main problem in this report; what are the
explicit forms for eigenvalues of row to row transfer matrices?
Some years ago, Reshetikhin[\analyR] conjectured it for the model 1
through the analytic
Bethe ansatz.
To present his result, we prepare some notations.
Let the vacuum state be such that all vertical edges assume $\epsilon_{-3}$.
By $\phi_{i}(u) (u=-3,\cdots, 3)$, we mean the eigenvalues, with
respect to this vacuum state, of the transfer
matrix of which the rightmost and the leftmost horizontal edges are assigned
edge variable $\epsilon_i$. Their explicit forms are given by,
$$\eqalignno{
\phi_{-3}(u) &= f(1+u)f(4+u)f(6+u), \qquad
\phi_{-2}(u) =\phi_{-1}(u) = f(u)f(4+u)f(6+u)  \cr
\phi_{0}(u) &=  f(u)f(3+u)f(6+u)  \qquad
\phi_{1}(u) =\phi_{2}(u) = f(u)f(2+u)f(6+u)  \cr
\phi_{3}(u) &=  f(u)f(2+u)f(5+u) \cr
f(u) &= \prod_{j=1}^N[u-w_j],       &{(\vacfund)}\cr
}$$
and $w_j (j=1, \cdots, N)$ are free parameters called inhomogeneity.
Note that we adopt different notations from those in [\analyR].
Let us further introduce two functions,
$$
D^{(1)}(u) = \prod_{j=1}^{N_1} [u- iu^{(1)}_j]  \qquad
D^{(2)}(u) = \prod_{j=1}^{N_2} [u- iu^{(2)}_j],
\eqno{(\dresses)}
$$
where the parameters $\{ u^{(a)}_j \}
(a=1,2 \quad j=1,\cdots, N_1 {\hbox{ or }}, N_2)$ are
solutions to the Bethe ansatz equations,
$$
{{f(iu^{(a)}_j+\delta_{a,p} {{s}\over{t_p}} )}\over
     {f(iu^{(a)}_j-\delta_{a,p}{{s}\over{t_p}} )} }
=\prod_{b=1}^{2}
{{D^{(b)}(iu^{(a)}_j+(\alpha_a|\alpha_b))}
\over{D^{(b)}(iu^{(a)}_j-(\alpha_a|\alpha_b))}} \qquad(a=1,2).
\eqno{(\bae)}
$$
Here $p, s$ specify the quantum space , now
$s=1$ and $p=2$.
The parameters $t_1$ and $ t_2$ are related to the lengths of
corresponding simple roots and $t_1=1, t_2=3$.
The numbers of roots $N_1, N_2$ are chosen such that
$N s\vL{p} -N_1 \alpha_1-N_2 \alpha_2$ is a non negative weight. \pn
Finally we introduce the most significant object in this report,
a Yangian analogue of the Young tableaux.
We associate an expression to a box with a number in the following
way:
$$\eqalignno{
\Fsquare(0.5cm, -3)  &\leftrightarrow
\phi_{-3}(u){{ D^{(2)}(u-1/2)}\over{ D^{(2)}(u+1/2)}}  \cr
\Fsquare(0.5cm, -2)  &\leftrightarrow \phi_{-2}(u)
      {{D^{(1)}(u-1) D^{(2)}(u+3/2)}\over{ D^{(1)}(u+2)D^{(2)}(u+1/2)}} \cr
\Fsquare(0.5cm, -1) &\leftrightarrow \phi_{-1}(u)
       {{D^{(1)}(u+5) D^{(2)}(u+3/2)}\over{ D^{(1)}(u+2)D^{(2)}(u+7/2)}}  \cr
\Fsquare(0.5cm, 0)   &\leftrightarrow \phi_0(u) {{D^{(2)}(u+9/2)
D^{(2)}(u+3/2)}\over
                            { D^{(2)}(u+5/2)D^{(2)}(u+7/2)}}  \cr
\Fsquare(0.5cm, 1)  &\leftrightarrow \phi_1(u)
     {{D^{(1)}(u+1) D^{(2)}(u+9/2)}\over{ D^{(1)}(u+4)D^{(2)}(u+5/2)}}  \cr
\Fsquare(0.5cm, 2)  &\leftrightarrow  \phi_2(u) {{D^{(1)}(u+7)
D^{(2)}(u+9/2)}\over
                              { D^{(1)}(u+4)D^{(2)}(u+11/2)}} \cr
\Fsquare(0.5cm, 3)  &\leftrightarrow
       \phi_3(u){{ D^{(2)}(u+13/2)}\over{ D^{(2)}(u+11/2)}}. \cr
&\phantom{  }   &{(\boxes)}
}$$
The  boxes in lhs are analogues to elements in
$V_{\vL{2}}$, $\epsilon_i, (i=-3,\cdots, 3)$.
We remark that these boxes carry spectral
parameter dependencies through rhs.\pn
These identifications stem from two reasons.
First, we identify the box of $i$  with the expression of which the vacuum
expectation value is $\phi_i(u)$.
Second, we make an analogy between the actions of Chevalley generators
 $E_{-\alpha_i}, i=1,2$
and the pole structure of expressions. For example, $\epsilon_{-3}$
and $\epsilon_{-2}$ are "connected" by the action of $E_{-\alpha_2}$.
On the other hand, we  see that the corresponding
expressions possess common poles
at $u=iu^{(2)}_j-1/2, (j=1,\cdots N_2)$ from eq.(\boxes).
The pole free condition
for the sum of these two is given by BAE. Then we regard that
these two expressions are "connected" by the pole free condition for
$\{u^{(2)}_j \}$.  In this way, we identify the "connection" by an
action of $E_{-\alpha_a}$ with that by the pole free condition for
$\{u^{(a)}_j \}$.\pn
We further remark the "crossing symmetry" like property of above boxes.
That is, we have a symmetry in box expressions,
$$
\Fsquare(0.5cm,i) = \Fsquare(0.5cm,-i)
|_{\scriptstyle u \rightarrow -6-u
     \atop \scriptstyle w_i \rightarrow -w_i }.
$$
This might be a consequence from the crossing symmetry for vertex weights.
At least, one can check that the configurations with the vacuum state
in its quantum space satisfy this property.\pn
Now the Reshetikhin's result can be neatly described in terms of the box
as,
$$
\tau^{(2)}_1(u) = \sum_{i=-3}^3 \Fsquare(0.5cm, i) .
\eqno{(\restau)}
$$
 Remark that we implicitly
assume that algebra acts on the auxiliary space.
Then the eigenvalues of two transfer matrices having
identical auxiliary space but different quantum ones will
have common combination of $D^{(a)}(u)$ functions, since
pole structure should be same. Indeed, this is the case for
$U_q(C_2)$ models[\AKC].\pn
In the following, we assume this pole free condition - $G_2$ action
correspondence for general cases.
The assumption is of great help in making conjecture for the eigenvalues
of transfer matrices of model 2,3,4.
We work out this program
 using weight space diagram for
classical $G_2$ algebra together with
BAE and the little information from
explicit fusion procedure.\pn
%
We first address the eigenvalue problem for the model 2.\pn
Let us  look at the 14 dimensional space $V_{\vL{1}}$ (Fig1).
A left-down(right-down) arrow presents the action of $E_{-\alpha_1}$
($E_{-\alpha_2}$).
 $W^{(1)}_1$ space has one extra vector having zero weight  other than
$v^{(\vL{1})}_7$ and $v^{(\vL{1})}_8$.
And they  are no longer identical to those in $V_{\vL{1}}$
as noted previously.
Nevertheless, we can gain lots of insight from Fig(1) following
the preceding hypothesis.
We succeed in making a conjecture for $\tau^{(1)}_1(u) $ using such arguments
, together with  the fact
$$
\tau^{(2)}_1(u-1/2) \tau^{(2)}_1(u+1/2) =\tau^{(1)}_1(u) +\cdots .
\eqno{(\tttot)}
$$
This is a natural consequence from the fusion procedure.\pn
The result can be again neatly described in terms of boxes.
For this purpose, we introduce a set of two boxes glued vertically.
We assign the spectral parameter $u+1/2 (u-1/2)$
to an upper (a lower) box .
Each column corresponds to the product of
two expressions with different spectral parameters.
Then $\tau^{(1)}_1(u)$ is given by
the sum of expressions corresponding to the following tableaux;\pn
\vskip 0.5cm
$$ \eqalignno{
&\Vtwobox(-3,-2) \qquad \Vtwobox(-3,-1) \qquad \Vtwobox(-3,0)
\qquad \Vtwobox(-3, 1) \qquad \Vtwobox(-3, 2) \qquad  \Vtwobox(-3,3)
\qquad \Vtwobox(-2,1) \qquad \Vtwobox(-2,2) \qquad \Vtwobox(-2,3)   \cr
&\Vtwobox(-1,1) \qquad \Vtwobox(-1,2) \qquad  \Vtwobox(-1,3)
\qquad \Vtwobox(0,3)   \qquad  \Vtwobox(1,3)   \qquad \Vtwobox(2,3) .   \cr
&\phantom{ } &(\twoboxes)
}$$
\vskip 0.5cm
\noindent(See Fig 2).
%
Next we consider the model 3, 4.
The model 3 possesses identical auxiliary space to the model 1.
By the argument soon below eq.(\restau),
 the problem reduces to evaluate vacuum expectation values.
They are explicitly calculable using a similar argument given
for $C_r$ models [\MacC].
 We omit details of
derivation and write only the modifications to $\phi_j(u)$\pn
model 3\pn
$$\eqalignno{
\phi_{-3}(u) &= \phi_{-2}(u) =f(1+u)f(5+u)  \cr
\phi_{-1}(u) &= \phi_{0}(u) =\phi_{1}(u) =f(1+u) f(11+u) \cr
\phi_{2}(u) &= \phi_{3}(u) =f(7+u)f(11+u) .
    &(\newvac)
}$$
%
We verified that
$\tau^{(2)}_1(u)$ is  pole free by virtue of the BAE(\bae) with $p=1$
and $s=1$.\pn
For the model 4, we can also evaluate the vacuum expectation values.
 Remarkably, the eigenvalue
for  model 4 transfer matrix obtained in this way agree with the
expression from  {\it the same
set of tableaux} in eq(\twoboxes), assuming $\phi_i(u)$ in eq.(\newvac), in
this turn.
This is naturally expected
from our arguments since  algebra acts on the auxiliary space.
And this coincidence supports validity of our hypothesis.\pn
 The expressions of $\tau$ for the model $1 \sim 4$ are checked
by comparison with results of
the brute force diagonalizations in $q \rightarrow 1$ limit
 for $N=2$  and some cases $N=3$
.\pn
%
%
\sj
\S 4 {\bf Solutions to the $T$-system} \pn
Before closing this report, let us consider the $T$-system problem.
The $T$-system is a set of functional relations among transfer matrices having
common quantum spaces,
however, different auxiliary spaces. For the $G_2$ model in the
present normalization of the spectral parameter, it reads[\KNSo]
$$\eqalignno{
T^{(1)}_m(u-3/2) T^{(1)}_m(u+3/2)
  &=T^{(1)}_{m+1}(u) T^{(1)}_{m-1}(u)  +g^{(1)}_m(u) T^{(2)}_{3m}(u) \cr
T^{(2)}_{3m}(u-1/2) T^{(2)}_{3m}(u+1/2)
   &=T^{(2)}_{3m+1}(u) T^{(2)}_{3m-1}(u)   \cr
   &+T^{(1)}_m(u-1/3)
     T^{(1)}_m(u)T^{(1)}_m(u+1/3)   \cr
T^{(2)}_{3m+1}(u-1/2) T^{(2)}_{3m+1}(u+1/2)
   &=T^{(2)}_{3m+2}(u) T^{(2)}_{3m}(u)   \cr
   &+T^{(1)}_m(u-1/2)
     T^{(1)}_m(u+1/2)T^{(1)}_{m+1}(u)   \cr
T^{(2)}_{3m+2}(u-1/2) T^{(2)}_{3m+2}(u+1/2)
   &=T^{(2)}_{3m+3}(u) T^{(2)}_{3m+1}(u)  \cr
   &+T^{(1)}_m(u)
     T^{(1)}_{m+1}(u-1/2)T^{(1)}_{m+1}(u+1/2)  \cr
&\phantom{  }   &(\tsys)
}$$
where
$g^{(1)}_m(u)=\prod_{j=1}^m g^{(1)}_1(u+{{3(m+1)}\over 2} -3j)$ and
$$
g^{(1)}_1(u)=
\cases{ f(-2+u) f(3+u) f(8+u)& $\hbox{ for }  p=2, s=1$ \cr
        f(-3/2+u) f(15/2 +u) & $\hbox{ for } p=1, s=1  $. \cr
}$$
And the same relations hold replacing $T$ by $\tau$ since
$T's$ constitute  a commuting family.
We can, in principle, obtain the explicit expression for
any $\tau^{(a)}_m(u)$ since now we have
expressions for $a=1,2 \quad m=1$.  \pn
In the limit $u \rightarrow \infty$, we expect that
 $T^{(a)}_m(u)$ converges to $Q^{(a)}_m$ after appropriate renormalization.
The latter gives  the dimension of Yangian space $W^{(a)}_m$
by specialization.
This means that the number of terms in $T^{(a)}_m(u)$ should
agree with the dimension of $W^{(a)}_m$.
The latter is known for small $m$.
Thus this deserves as a check.
By successive solving eq.(\tsys),
we obtain  $T^{(1)}_1(u), T^{(2)}_m(u) (m=1,2)$,  consisting of
34,133,92 terms respectively, which
agree with $Q^{(a)}_m$[\AKBAE].\pn
In particular, the expression for  $Q^{(1)}_m$ is conjectured
in a beautiful form [\AKBAE],
$$
Q^{(1)}_m =\sum_{j=0}^m \chi(j\Lambda_1)
\eqno{(\yangch)}
$$
where $\chi(\Lambda)$ is a classical $G_2$ character for
the highest weight module $V_{\Lambda}$.\pn
Thus it might be tempting to find a "u-" version of this.
Through studies on the $T$-system, we find a conjecture for
the eigenvalue of transfer matrix $T^{(1)}_m(u)$.
To express this, we prepare a set of tableaux $BW^{(1)}_m$.\pn
{\it Definition}\pn
Let $b(m, \{u\}, \{d\})$ be a table
consisting of 2$\times$ m boxes and whose upper
(lower) row arguments are $\{ u_1,u_2, \cdots, u_m \},
(\{ d_1,d_2, \cdots, d_m \})$ where $-3 \le u_i, d_i \le 3$.
$$
	\normalbaselines\m@th\offinterlineskip
	\vcenter{
   \hbox{\Fsquare(0.5cm,u_1)\naga \hskip-0.4pt\Fsquare(0.5cm,u_m)}
	      \vskip-0.4pt
   \hbox{\Fsquare(0.5cm,d_1)\naga\hskip-0.4pt\Fsquare(0.5cm,d_m)}
    }
$$
Then $b(m, \{u\}, \{d\})$ is a member of $BW^{(1)}_m$
if it satisfies
\item{1.} Each column is a member of tableaux in eq.(\twoboxes).
\item{2.} Each adjacent columns should satisfy the condition,
\itemitem{2.a} $ u_k \le u_{k+1}, d_k \le d_{k+1} \quad (k=1,\cdots,m-1) $.
\itemitem{2.b} if $u_k=u_{k+1}=-3$ then either of $d_k$ or $d_{k+1}$ must
     be less than $0$.
\itemitem{2.c} if $d_k=d_{k+1}=3$ then either of $d_k$ or $d_{k+1}$ must be
           greater than $0$. \pn
{}From $b(m, \{u\}, \{d\}) \in BW^{(1)}_m$,
we construct an  expression $b(m, \{u\}, \{d\}, u) $
in the following way,
\item{1.} For a box with figure $u_k, (d_k)$ we assign the spectral
parameter $u-{{3(m+1)}\over 2}+3k+{1\over 2},
(u-{{3(m+1)}\over 2}+3k-{1\over 2})$
\item{2.} Take the all products of the corresponding expressions
 according to expressions in (boxes) and with the above mentioned spectral
parameters.\pn
Then we have ,\pn
\vskip 0.5cm
\noindent{\bf Conjecture}.\pn
The eigenvalue $\tau^{(1)}_m(u)$ of $T^{(1)}_m(u)$ is given by
$$
\tau^{(1)}_m(u)= \sum b(m,\{u\}, \{d\}, u)
\eqno{(\soltau)}
$$
where the summation should be taken over all
$b(m, \{u\}, \{d\}) \in BW^{(1)}_m$. \pn
We have the following supports to this conjecture.
\vskip 0.5cm
\item{1} We have verified that the number of elements in $BW^{(1)}_m$
agree with the Yangian dimension up to $m=10$ (dimension $=73788$).
\item{2} The expression for $T^{(1)}_2(u)$  is analytically verified
 to obey the above rule for $p=1,2$ and $s=1$.
\item{3}  If  $\tau^{(1)}_m(u)$ defined by eq(\soltau) is really the
solution to the $T$-system(\tsys), then this equality must hold good
irrespective
of forms for $f(u), D^{(1)}(u)$ and $ D^{(2)}(u)$.
We assume 3rd order polynomials in $u$ for $f(u), D^{(1)}(u),  D^{(2)}(u)$
, and  choose their coefficients from random numbers.
Then we have checked the agreement between
the numerical solution for the $T$-system  and the numerical result
from the above rules up to $m=6$.\pn
\sj
\S 5 {\bf Conclusion}\pn
In this paper, we have reported studies on fusion $U_q(G_2)$ vertex models.
Under few assumptions, we have found analytic expressions for
the transfer matrices' eigenvalues of models related to
$W^{(1)}_1$ or $W^{(2)}_1$.
It has been shown how the result can be neatly expressed in terms
of the tableaux.
A conjecture for $\tau^{(1)}_m(u)$ is given and several supports
for this are presented. We have seen that the rules for tableaux
are quite simple.\pn
These facts encourage us to extend similar analyses on models
based on other Lie algebras. Such program is now under
progress and partial results are promising[\AKJS].\pn
Contrary to the success for $a=1$ case,
we still have not yet determined general expressions for $\tau^{(2)}_m(u)
\quad (m=4,\cdots \infty)$. This is partly due to the lack of
information even for $Q^{(2)}_m$[\AKBAE].
We hope to report the further studies on this in near future.\pn
\sj
\noindent{\bf Acknowledgment}\pn
The author would like to thank A. Kuniba and T. Nakanishi for discussions,
A. Kuniba for comments and critical reading of the manuscript.
He also thanks to  T. Nakashima for allowing him to use tex macro
produced by   K. Nakahara.
\vfill\eject
%
%
\noindent{\bf References}\pn
\item{\Yang} C.N. Yang, Phys. Rev. Lett. {\bf 19} (1967) 1312.
\item{\Bax} R.J. Baxter, {\sl Exactly Solved Models in Statistical
  Mechanics} (1982) (London, Academic Press).
\item{\Jimbo} M. Jimbo, Lett.Math.Phys. {\bf 10} (1985) 63.
\item{\Baz} V.V. Bazhanov, Phys.Lett. B. {\bf 159} (1985) 321.
\item{\KSR} P.P. Kulish, E.K. Skylanin and N.Yu. Reshetikhin,
             Lett. Math. Phys. {\bf 5} (1981) 393.
\item{\OW} E.I. Ogievetsky and P.B. Wiegmann, Phys.Lett. B.
     {\bf 189} (1987) 125.
\item{\RW} N.Yu. Reshetikhin and P.B. Wiegmann, Phys.Lett. B.
     {\bf 168} (1986) 125.
\item{\KirR} A.N. Kirillov and N.Yu. Reshetikhin, J.Sov. Math.{\bf 52}
            (1990) 3156.
\item{\Drin} For quantum affine Lie algebra, see  [{\Jimbo}] and
    V.G. Drinfel'd, {\sl Proceedings of ICM} Berkley 1987
  (Providence, RI: American Mathematical Society).
\item{\ChPrcmp}  V. Chari and A. Pressley, Comm. Math. Phys. {\bf 142}
(1991) 261.
\item{\KNSo} A. Kuniba, T. Nakanishi and J. Suzuki,
         "Functional relations in solvable lattice models I"(hep-th.9309137),
         Int.J.Mod.Phys.A, to be published .
\item{\AKC}  A. Kuniba, J.Phys. A{\bf 27}(1994) L113.
\item{\Resan} N.Yu. Reshetikhin, Sov. Phys. JETP {\bf 57}(1982) 691.
\item{\BR} V.V. Bazhanov and N.Yu. Reshetikhin, J.Phys. A{\bf 23}(1994) 1477.
\item{\Og} E.I. Ogievetsky, J.Phys. G{\bf 12}(1986) L105.
\item{\AKG} A. Kuniba, J.Phys. A{\bf 23}(1990) 1349.
\item{\KSG} A. Kuniba and J. Suzuki, Phys. Lett. A.{\bf 160}(1991) 216.
\item{\ChPra} V. Chari and A. Pressley, J. Reine. Angew. Math.{\bf 417} (1991)
87.
\item{\analyR} N.Yu. Reshetikhin, Lett. Math. Phys. {\bf 14}(1987) 235.
\item{\MacC} N.J. MacKay, Nucl. Phys. {\bf B356}(1991) 729.
\item{\AKBAE}  A. Kuniba, Nucl. Phys. {\bf B389}(1993) 209.
\item{\AKJS} A. Kuniba and J. Suzuki, in preparation.
%
%
\vfill\eject
\noindent{\bf Figure Captions}\pn
Figure 1. 14 dimensional space for $V_{\vL{1}}$.
The numbers associated with boxes are assigned according to the
indices of vectors $\in V_{\vL{1}}$ given in ref(\AKG). \pn
Figure 2. 15 dimensional space for $W^{(1)}_1$.
Boxes with $1 \le i \le 8$ denote $v^{\vL{1}}_i$ in [\AKG].
Those with $9 \le i \le 14$ represent $v^{\vL{1}}_{i-8}$ by replacing
$v_{\mu}$ by $v_{-\mu}$.
\bye